\documentclass[aps,prd,reprint,showpacs]{revtex4-1}

\usepackage{dsfont}
\usepackage{dcolumn}
\usepackage{amsmath}
\usepackage{amssymb}
\usepackage{physics}
\usepackage{float}
\usepackage{graphicx}
\usepackage{bm}
\usepackage[colorlinks = true, linkcolor = blue, urlcolor = blue, citecolor = red, anchorcolor = blue]{hyperref}
\usepackage{verbatim}
\usepackage{xcolor} 
\usepackage{soul} 
\usepackage{enumitem}

\begin{document}
\title{Constraining the number of fundamental quantum degrees of freedom using gravity}
\author{ Harshit Verma}
 \email{h.verma@uq.edu.au}
\affiliation{Australian Research Council Centre of Excellence for Engineered Quantum Systems (EQUS), School of Mathematics and Physics, The University of Queensland, St Lucia, QLD 4072, Australia}

\author{Magdalena Zych}
\affiliation{Australian Research Council Centre of Excellence for Engineered Quantum Systems (EQUS), School of Mathematics and Physics, The University of Queensland, St Lucia, QLD 4072, Australia}

\author{Fabio Costa}
\email{f.costa@uq.edu.au}
\affiliation{Australian Research Council Centre of Excellence for Engineered Quantum Systems (EQUS), School of Mathematics and Physics, The University of Queensland, St Lucia, QLD 4072, Australia}
\date{\today}
\begin{abstract}
We consider the effect of gravity on extended quantum systems (EQS) in the low energy regime. We model the gravitational effect due to a nearby source mass as a redshift in the internal Hamiltonian of the EQS. Due to the dependence of the energy spectrum of the EQS on the position of the massive particle (via the redshift) at zero temperature, our model predicts gravitational decoherence of the massive particle in the position basis. We show that the decoherence effect is multiplicative, in the sense that the increase in number of EQS gravitationally interacting with a single massive particle leads to an increase in its decoherence. If the considered model of gravitational redshift holds alongside the linearity of quantum mechanics (as an appropriate limit of an accepted theory for coupling quantum matter with gravity) to allow for a spatial superposition of the source mass, we propose that the same methodology can be applied to continuous fields, which are essentially EQS. This could provide an upper limit on the number of undiscovered fields by observing coherent superpositions of masses, e.g., in a matter wave interferometer. Besides, by taking a spin chain as a toy model of an EQS, we analyse the dependence of the new effect on relevant system parameters and identify the number of independent spin chains that can cause a detectable effect.
\end{abstract}
\maketitle

\section{Introduction}

One of the greatest successes of theoretical physics has been in reducing an incredibly vast range of phenomena to the interactions between a relatively small number of fundamental degrees of freedom (DOF), summarised by the Standard Model of particle physics \cite{Mann2010}.

Despite its undisputed experimental success, it is generally believed that the Standard Model is incomplete, not least for the fact that it does not include gravity. The common understanding is that new physics should emerge at sufficiently high energy scales, in the form of new particles. String Theory, for example, implies the existence of an infinite number of particles, emerging as a tower of excitations of an infinite number of fundamental DOF \cite{Polchinski1998, Green2012}. Although theoretical arguments have been proposed to bound the maximum number of DOF in a fundamental theory  \cite{Witten2001, Anninos2019, Bousso2002}, it is generally believed that present-day experiments impose no constraints: particles of sufficiently large mass are assumed to have no observable effect on low-energy experiments.

Here we propose a mechanism through which a sufficiently large number of local DOF would reduce the coherence in low-energy interference experiments. The mechanism relies on the universal coupling due to general relativistic redshift \citep{Zych2011, Zych_2012, pikovski2015, Zych2016, pikovski2017}, which has been the subject of increased interest in recent years in the broader context of proposed low-energy tests of the interplay between quantum theory and gravity \citep{Bose, Carney_2019, Vedral, singleE, singleP, Markus, Bose2, Guerreiro_2020, unknown, Verma2020, Schwartz_2019, Zych_2018}.

\begin{figure}[ht!]
  \begin{center}
    \includegraphics[trim={8cm 0.5cm 11cm 0cm},clip,width=0.9\columnwidth]{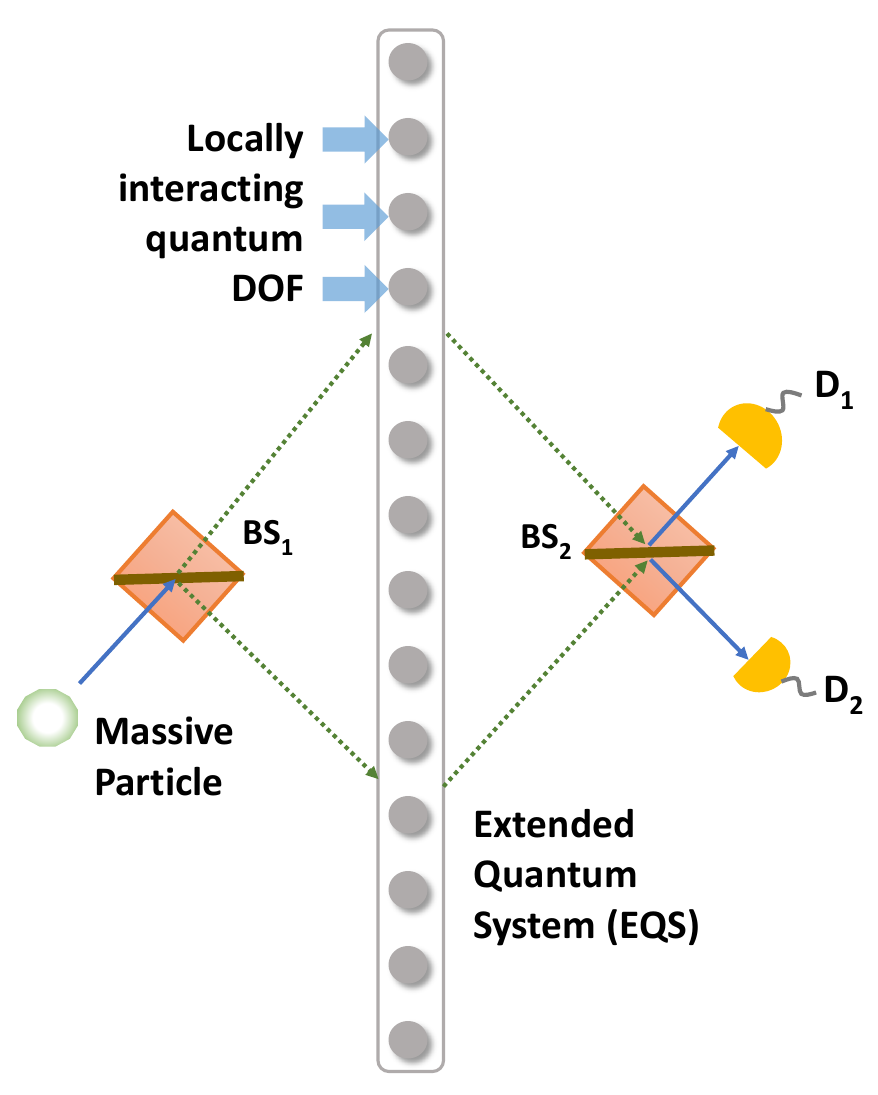}
  \end{center}
 \caption{Schematic diagram of the \textit{Gedanken} experiment. A massive particle passes through a Mach-Zehnder interferometer in the vicinity of an extended quantum system (EQS). Due to the redshift caused by the massive particle, the EQS has a different ground state depending on the particle's path, at zero temperature. A superposition of paths results in entanglement between the EQS and the particle, reducing the latter's coherence. $BS_1, BS_2$ denote beams-splitters, while $D_1$, $D_2$ are detectors. }
  \label{fig:ged_chain}
\end{figure}

In the typical setting studied so far, one considers a particle with internal DOF that effectively act as a ``clock'', namely they evolve at a rate given by the particle trajectory's proper time. By preparing the particle in a superposition of two different heights in a gravitational potential (or, more generally, a superposition of two trajectories with different proper times), each amplitude evolves by a different amount, with the internal states encoding which-way information and thus reducing the visibility in an interference experiment \cite{complement}. In contrast, here we consider the dynamical DOF to be in a fixed location, while the source of the gravitational field (a massive particle) is prepared in a superposition. We take as an example an EQS in the vicinity of a massive-particle interferometer, as shown in Fig.~\ref{fig:ged_chain}. 
For each arm of the interferometer, the particle induces a gravitational redshift on a different section of the EQS, resulting in entanglement between the quantum DOF constituting the EQS and the position of the particle. This entanglement, thus, leads to a loss in the interferometric visibility in an interference experiment with the massive particle. It is important to note here that though a quantum superposition of a massive particle has been demonstrated before experimentally \cite{Asenbaum2017a}, and forms the backbone of many other proposals (see for example \cite{pikovski2015, pikovski2017, Bose, Bose2}), one can not rule out the presence of collapse mechanisms driven by gravitational effects \citep{Toros2018}, which could deem the premise unfeasible beyond (as yet undetermined) mass scale. Therefore, our model is speculative, in the sense that it assumes that there is no fundamental barrier in treating the spatial DOF of a massive particle (source mass for gravitational effects) as a quantum DOF, and hence, belongs to the realm of untested physics.

Although the above effect of decoherence is interesting in its own right, it is too small to be relevant for experimentally controllable systems and realistic matter-wave interferometers, presently available. However, since the effect is generally valid for all EQS, we can consider a toy model to illustrate it. To this end, we take an example of quantum spin chains as EQS, and consider the decoherence offered by a single spin chain (as a proxy for a single field DOF). Multiple spin chains can then mimic multiple field DOF and thus also multiple fundamental particles. We find that, for a sufficiently large number of chains, the cumulative effect becomes arbitrarily large, eventually producing a measurable loss of interferometric visibility. The implication is that present-day matter-wave interference experiments would not be possible given an arbitrarily large number of background DOF (quantum fields). Although our analysis does not provide a concrete number, it indicates that interference experiments could impose an upper bound on the number of elementary particles. It is worth highlighting that the effect, and the decoherence mechanism that we propose, is valid in principle for all masses , including present day matter-wave interferometry. For the  relatively small value of mass that the current interferometers are able to superpose, the bound so obtained may still permit a very large number of local DOF. In order to derive ab-initio the expected magnitude of the decoherence effect one would need detailed modelling of both: the specific matter-wave interferometer as well as the fundamental theory containing a large number of local DOF (e.g. quantum field theory beyond the Standard Model). Nonetheless, the fact that present-day experiments could provide \textit{any} bound would still be highly significant, as it would rule out models with infinitely many types of particles.

The manuscript is organized in the following manner: In Sec.~\ref{framework}, we discuss the differential redshift of an EQS Hamiltonian caused by a nearby massive particle. We then establish the novel phenomenon of gravitational decoherence of a massive particle in a spatial superposition caused by such redshift. We show that the strength of this decoherence can be used to estimate the number of EQS interacting with the particle. In Sec.~\ref{schain}, as an illustrative example of EQS, we consider a spin chain. We calculate the redshift numerically and demonstrate the gravitational decoherence phenomenon, while also identifying the factors that could affect the coherence in a proof-of-principle experiment with spin chains. Moreover, we also consider a limiting case with a very high mass of the particle to provide an upper bound on the number of spin chains required to cause a significant decoherence. In Sec.~\ref{discuss}, we discuss relaxing the  assumptions in the Gedanken experiment and argue that the decoherence effect would still remain. In Sec.~\ref{conc}, we list the significant conclusions from the manuscript.

\section{Theoretical Framework} \label{framework}
For simplicity, we consider a generic one-dimensional Hamiltonian, which describes many relevant EQS. For such systems, particularly those with local interactions, the Hamiltonian can be written as a sum of local Hamiltonians as shown below:
\begin{eqnarray} 
H_0=\sum_{i=1}^L h_i~,
\label{eq:1dlocalH}
\end{eqnarray}
where it is assumed that the system is composed of L subsystems and where each $h_i$ acts non-trivially only on a set of ``neighbouring'' quantum DOF. Therefore, such Hamiltonians are composed of several local Hamiltonians supported on distinct local Hilbert spaces. Such  \textit{tensored} structured Hamiltonians typically describe physical EQS such as spin chains, and notably also the discretized forms of quantum fields.

We now describe the gravitational red-shift effect arising in an EQS due to a gravitating particle nearby, while assuming the Hamiltonian to be of the general form in Eq.~\eqref{eq:1dlocalH}. This will allow us to establish the salient qualitative features of the novel gravitational decoherence phenomenon in the later part of this section.

We consider a scenario wherein a massive particle is located in proximity to the EQS such that the two systems interact gravitationally. Our model assumes a fixed position of the EQS and looks at the gravitational redshift in its Hamiltonian, while it is fixed in space. 

Such a gravitational effect can be interpreted as differential redshift along the extended system. The redshift factor is non-uniform along the system as it depends on the relative position of the massive particle and the subsystem of the EQS labelled by ``i". This redshift effect can be expressed as: $h_i\rightarrow \sqrt{-g_{00}(x_i)}h_i$, where $g_{00}$ is the time-time component of the metric sourced by the massive particle (with signature $(-,+,+,+)$). Here, $x_i$ the position of the $i^\mathrm{ th}$ subsystem with the local Hamiltonian $h_i$, with respect to the massive particle. Here we assume that $g_{00}$ does not depend on time, i.e., a static spacetime background. As we associate with each subsystem $i$ their position $x_i$, we implicitly assume that the subsystem does not have a significant spatial span on the length scale on which $g_{00}$ appreciably varies.
Therefore, the total Hamiltonian of the EQS, including the gravitational effect of differential redshift is given as follows:
\begin{eqnarray}
H_\textrm{total}= \sum_{i=1}^L  \sqrt{-g_{00}(x_i)}h_i~.
\label{eq:ham_gen}
\end{eqnarray}
We stress that, so far, $x_i$ is just the classical relative distance and the relevant metric component $g_{00}$ is well defined classically. 
In general relativity, as a direct consequence of the Einstein Equivalence principle (EEP), one obtains a red-shift in the energy of a test system due to the local distribution of mass-energy (which essentially defines the metric). This is termed as universality of gravitational redshift (UGR) \cite{Will2014}. The UGR guarantees that the treatment of gravitational effect on the Hamiltonian in Eq.~\eqref{eq:1dlocalH}, represented in Eq.~\eqref{eq:ham_gen} remains valid for any physical realisation of the EQS.

In this scenario, if one considers the temperature of EQS to be zero ($T=0$), then it would always remain in the ground state (the state with lowest energy) irrespective but not independent of the position of the massive particle. This proposition relies on the assumption that the massive particle is brought close to the EQS adiabatically such that the system is allowed to relax to the ground state of the instantaneous Hamiltonian,

If the massive particle is in a positional state $|\tilde{r}\rangle$, the state of the massive particle and the EQS is given as follows:
\begin{eqnarray}
|\psi\rangle = |\tilde{r}\rangle~|GS_{\tilde{r}} \rangle~,
\label{eq:ent}
\end{eqnarray}
where $|\psi\rangle$ is the state of the complete system and $|GS_{\tilde{r}}\rangle$ represents the ground state of the EQS (i.e., of $H_\textrm{total}$) when the gravitating particle is at a particular position labelled by $\tilde{r}$ which also represents its positional state.

We now apply the above reasoning to the scenario when a massive particle is in spatial superposition. Though the extension of Eq.~\eqref{eq:ent} to this scenario seems straightforward, it rests on some assumptions. Note that the position of the massive particle is a quantum degree of freedom, and hence, the Hamiltonian for the total system would not be compatible with a single, classical spacetime metric. The validity of such a scenario has been widely debated in the light of proposed Di\'osi-Penrose collapse which hypothesizes that such superpositions of spacetime may not be possible entirely \cite{Diosi1989, Penrose1996}. Though the experimental test of such collapse models in the complete range of relevant parameters is still ongoing \cite{Donadi2021, Toros2018}, our proposal assumes that the source mass can indeed be prepared in a spatial superposition, in analogy to the test masses in existent matter-wave interferometry experiments where particles of substantial (though relatively small) mass have been successfully demonstrated to be in a spatial superposition \cite{Asenbaum2017a}.

Therefore, we consider a \textit{Gedanken} experiment wherein the interferometric setup in Fig.~\ref{fig:ged_chain} represents the underlying physical setting. At first, the beamsplitter $BS_1$ prepares the massive particle in a spatial superposition $|+_{\tilde{r}} \rangle = \frac{1}{\sqrt{2}}\big[|\tilde{r}_1\rangle + |\tilde{r}_2\rangle \big]$, where $|\tilde{r}_1\rangle, |\tilde{r}_2\rangle$ are two positional states near the EQS, which are assumed to be orthogonal. If the spatial superposition is prepared sufficiently slowly for each amplitude of the spatial DOF, the EQS will remain in the ground state associated with the particle's position, leading to the following entangled state:
\begin{eqnarray}
|\psi\rangle =\frac{1}{\sqrt{2}} \big[|\tilde{r}_1\rangle~ |GS_{\tilde{r}_1} \rangle~+|\tilde{r}_2\rangle~ |GS_{\tilde{r}_2} \rangle\big]~.
\label{eq:GSpos}
\end{eqnarray}

To reiterate, we resort to regime of low energy, which allows us to incorporate the effect of GR as a redshift in the Hamiltonian of a quantum system, without requiring explicit quantization of the metric. \footnote{A simplified version of the required Hamiltonian, sufficient to account for the positional states of the massive particle involved in the experiment, is {$\sum_{i=1}^L  \sum_{j=1}^2\sqrt{-g^j_{00}(x_i)} |\tilde{r}_j\rangle \langle \tilde{r}_j| \otimes h_i$, where $g^j_{00}(x_i)\equiv g_{00}(x_i-\tilde{r}_j)$}
denotes the metric associated with the massive particle at position $\tilde{r}_j$}. Indeed, no dynamical DOF associated with the metric participates in the effect; the only role of the metric is to mediate an interaction between the massive particle and the EQS. This is fully analogous to the interaction of two particles through a potential, such as gravity in the Newtonian limit or the Coulomb interaction between charges, which do not require quantization of the gravitational or electromagnetic fields respectively. Therefore, within the physical regime under consideration, Eq.~\eqref{eq:GSpos} relies uniquely on the validity of the classical redshift and on the linearity of quantum mechanics, independently of any detail of a  theory of quantum gravity. The approach is valid as long as the dynamics of the gravitational DOF is not relevant, i.e. the timescale of the creation of the spatial superposition is slow \cite{Markus, Mari2016}. 

Furthermore, quantum states can be defined on space-like surfaces associated with a suitable global time coordinate, which can be interpreted as the time measured by an asymptotic observer, free from the effect of variations in the local mass-energy density stemming from the spatial superposition of mass. Again, this is fully analogous to the definition of global coordinates in the Newtonian limit, where gravity acts as a potential in a fixed spacetime. See, e.g., Ref.~\cite{Zych2019} for further details on the formalization of general relativistic effects arising from the superposition of masses. 

Since the ground state of the EQS is entangled with the position of the massive particle, this will lead to decoherence of spatial superposition of the massive particle. The overlap between the ground states of EQS corresponding to different positions of the massive particle is given as follows:
\begin{eqnarray}\label{visibility}
\mathcal{M} = |\langle GS_{\tilde{r}_1}|GS_{\tilde{r}_2}\rangle|~.
\end{eqnarray}

\begin{figure}
  \begin{center}
    \includegraphics[trim={2.5cm 4cm 10cm 4cm},clip,width=1\columnwidth]{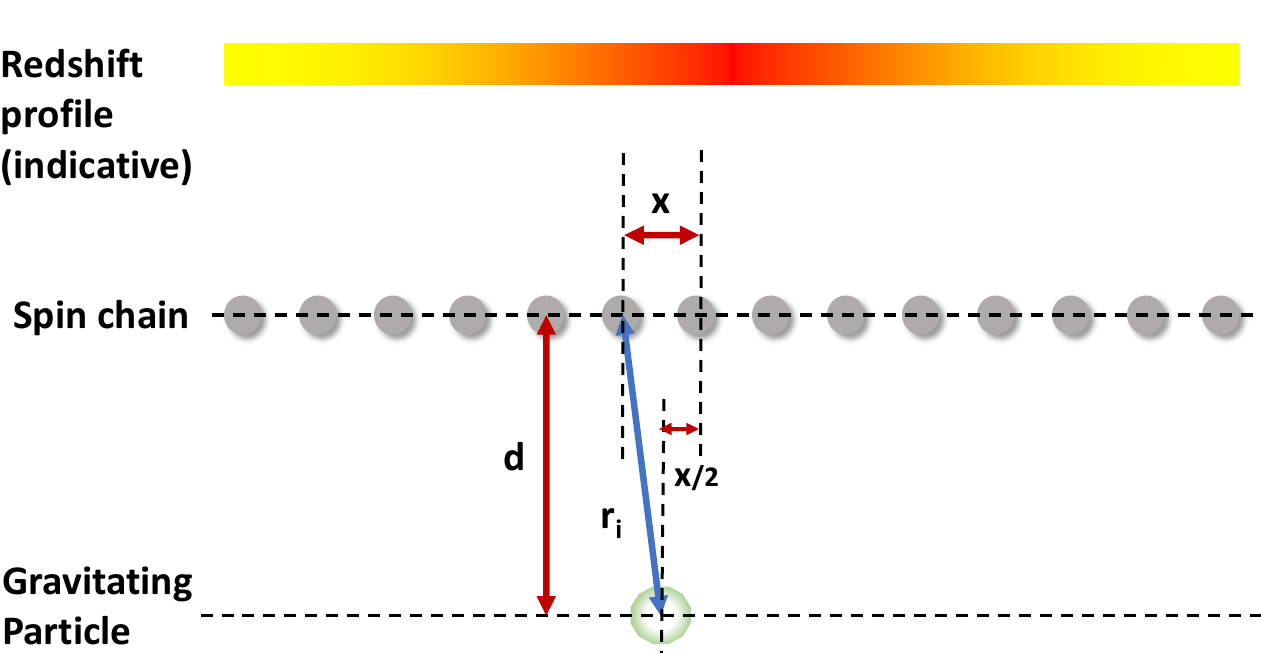}
  \end{center}
 \caption{Schematic diagram of the proposal with system parameters denoted. The massive particle is located at a perpendicular distance of d from the chain. x denotes the lattice spacing within the chain.}
  \label{fig:grav}
\end{figure}

In our Gedanken experiment, the superposed amplitudes of the mass are further overlapped at the second beamsplitter, see Fig. \ref{fig:grav}. In that context, $\mathcal{M}$ is equal to the interferometric visibility, which is experimentally obtained from the probabilities of detection of the particle at the outputs of the interferometer (at detectors $D_1$ and $D_2$ in
Fig. 1). Such a detection corresponds to the measurement in the basis $|\pm_{\tilde{r}} \rangle = \frac{1}{\sqrt{2}}\big[|\tilde{r}_1\rangle \pm |\tilde{r}_2\rangle \big]$. The overlap $\mathcal{M}$ also
quantifies the distinguishability between the two involved
states \cite{complement}, which here is $\mathcal{D} = \sqrt{1-\mathcal{M}^2}$, and represents the coherence of the spatial superposition of the mass.  Therefore, a low value of $\mathcal{M}$ signifies a higher distinguishability, signalling the decoherence of the spatial superposition of the interfering massive particle. Here, we nominally call ($1-\mathcal{M}$) as the distinguishability, which also agrees with the complementarity principle mentioned above.

In general, we expect the effect to be extremely small: the gravitational interaction of a particle with the EQS will result in only slightly different ground states, so that $\mathcal{M}$ will be close to one.  In this regime, the nominal definition matches the actual one in the limit when visibility is quite close to one. Crucially, we are going to see that, in general, $\mathcal{M}$
will be strictly less than 1, which means that if more extended
systems are present their effects will accumulate
and the decrease in the coherence of the massive superposition
will be stronger.

\subsection{Decoherence caused by many independent EQS}
Let us consider a scenario with $\mathcal{N}$ independent EQS, each interacting independently with the massive particle, so that their joint ground state has the form $\ket{GS_{\tilde{r}}}^{\otimes \mathcal{N}}$. If each system contributes $\approx \mathcal{M}$ decoherence, the combined system will contribute $\left| (\bra{GS_{\tilde{r}}}^{\otimes \mathcal{N}})(\ket{GS_{\tilde{r}}}^{\otimes \mathcal{N}})\right| \approx \mathcal{M}^{\mathcal{N}}$. As long as $\mathcal{M}<1$, we have $\mathcal{M} ^ \mathcal{N} \rightarrow 0$ for sufficiently large $\mathcal{N}$. In other words, no matter how small the effect, there will be some $\mathcal{N}$ large enough to produce significant decoherence. This means that experimental observation of quantum interference can, in principle, put an upper bound on the total number of DOF with which the particle interacts gravitationally.

Fundamentally, such gravitationally induced decoherence can also be considered with continuous EQS such as a field. The basic argument remains the same: a local Hamiltonian has the form $H=\int dx\, h(x)$, which, under the effect of gravitational redshift, becomes $H_{\textrm{total}}=\int dx \sqrt{-g_{00}(x)}h(x)$. As before, the ground state $|GS_{\tilde{r}}\rangle$ of the red-shifted Hamiltonian $H_{\textrm{total}}$ depends on the position $\tilde{r}$ of the massive particle, and the interferometric visibility is given by the overlap between ground states corresponding to different positions of the particle, Eq.~\eqref{visibility}. As the two ground states are different, we expect their overlap will be in general strictly less than one.

Crucially, a model comprising different particle species---hence, different fields---has a state space given by the tensor product of all the fields. If the interactions between the fields are weak as compared to the masses of the corresponding particles, the total ground state will be, to a first approximation in perturbation theory, a product of the individual ground states of the different fields, $\ket{GS_{\tilde{r}}}\approx \bigotimes_j \ket{GS^j_{\tilde{r}}}$. Therefore, the overlap between two ground states, corresponding to the massive particle at positions $\tilde{r}_1$, $\tilde{r}_2$, is the product of the overlaps: $\braket{GS_{\tilde{r}_1}}{GS_{\tilde{r}_2}}\approx \prod_j \braket{GS^j_{\tilde{r}_1}}{GS^j_{\tilde{r}_2}} $.  Finally, because of the universality of the gravitational redshift, we expect the ground-state overlap for each field type to be of a similar order of magnitude, $\braket{GS^j_{\tilde{r}_1}}{GS^j_{\tilde{r}_2}}\approx \mathcal{M}$, leading to the prediction that the visibility is exponentially suppressed in the limit of arbitrarily many fields.

According to the argument above, interference experiments with massive particles have an upper bound on the achievable visibility. Although such a bound would be very close to one in a theory with a limited number of fundamental particles (such as the standard model), it becomes significantly smaller when the number of fundamental particles is assumed to be large. This is how the spatial coherence of the massive particle observed in an actual interference experiment can provide an estimate on the number of independent quantum fields and hence, the number of extant fundamental particles.

\section{Numerical results with spin chain as a toy model for EQS} \label{schain}

In this section, we devise a simple, concrete realisation of our Gedanken experiment where the EQS is modeled as a spin chain and there is a massive particle of mass $M$ situated at a distance $d$ from the chain. We denote the chain’s lattice spacing as $x$ and consider that $x\ll d$,
see Fig.~\ref{fig:grav}. We denote by $L$ the length of the spin chain, i.e. the total number of sites.  Fig.~\ref{fig:grav} also includes a ‘redshift profile’,
which represents the differential redshift in the spin chain
Hamiltonian due to gravitational potential of the massive
particle nearby. Later, we use this model to quantify the coherence of the spatial superposition of the massive particle. We also note that spin chains as a toy model EQS enable the possibility of a proof of principle experiment by simulation of the gravitational effects on quantum hardware \citep{Simon2011, I_trap}.

We consider here an XXZ type chain with open boundary conditions whose Hamiltonian is given as follows:
\begin{eqnarray}
H &=& J_x \sum_{i=1}^{L-1} \big(\hat{S}_i^x\hat{S}_{i+1}^{x} +\hat{S}_i^y\hat{S}_{i+1}^{y}\big) \nonumber\\
&+& J_z\sum_{i=1}^{L-1} \hat{S}_i^z\hat{S}_{i+1}^{z} - \sum_{i=1}^L B_i \hat{S}_i^z~,
\label{eq:hamil}
\end{eqnarray}
 where $\hat{S}_i$ represents the spin operator at the site $i$ of the spin chain, $J_x, J_z$ are the nearest neighbour coupling (interaction constants) and $B_i$ is the magnetic field at the site labelled $i$. As a special case, we also consider the XX Hamiltonian (setting $J_z=0$ in Eq.~\eqref{eq:hamil}) which can be mapped to a free fermionic system through the Jordan-Wigner transformation \cite{LIEB1961407}. This model represents a scenario of fundamental interest wherein a massive particle interacts gravitationally with a fermionic field.

For a spin chain Hamiltonian which consists of various local interaction terms and can be written as a sum of local Hamiltonians, i.e.\ $H=\sum_i h_i$, to the first order in relativistic correction (and taking $c=1$), the effect of a nearby massive particle is given as follows:
\begin{eqnarray}
H_\textrm{total}=\sum_i h_i (1+\phi(\Vec{r_i}))~,
\label{eq:hamm}
\end{eqnarray}
where $\phi(\Vec{r_i}) = -G\frac{M}{|\Vec{r_i}|}$, which is the Newtonian potential, $\Vec{r_i}$ is the displacement vector from the particle to the site $i$ in the spin chain and $G$ is the universal gravitational constant. The above expression for the redshifted Hamiltonian stems from expanding the terms of $\sqrt{-g_{00}(x_i)}$ in Eq.~\eqref{eq:ham_gen} in the post Newtonian approximation which gives: $-g_{00} \approx 1+2\phi$ , and thus, $\sqrt{-g_{00}}\approx 1+\phi$. We shall rely on this form of the total Hamiltonian in the later sections to calculate the ground states corresponding to different positions of the massive particle and subsequently all the quantities of our interest, including the overlap.

\subsection{Modelling differential redshift in spin chain}

As a simple model, we consider that the particle is at a distance $d$, and between the sites $k$ and $k+1$ of the spin chain, see Fig.~\ref{fig:grav}. Recall that to incorporate the redshift effect described in Eq.~\eqref{eq:displacement}, we have to associate each term of the Hamiltonian to a position along the chain. We see that there are two different types of terms: the single-site terms, $B_i \hat{S}_i^z$ which can be taken to be located on the site $i$; and terms describing the coupling energy
between neighbouring sites. As a reasonable approximation, we associate the latter type to the middle point between the sites. In the limit of a large number of sites, i.e. $L$ is large, the details of where with respect to the sites we associate the energy terms should disappear. However, for numerical purposes, this choice appears to produce the most stable results.
This modelling means that we consider the single site term in Eq.~\eqref{eq:hamil} i.e. $\sum_{i=1}^L B_i \hat{S}_i^z$ to have a different redshift factor than the spin-spin interaction terms owing to the geometry of the particle placement shown in Fig.~\ref{fig:grav}. Hence, in the redshift factors: $\phi_i (r_i)$ and $\phi_{i,i+1} (r_i)$ for single site and the spin interaction terms respectively, the displacement vectors $\vec{r}_i$ are:
\begin{eqnarray}
r_i &=& \sqrt{\big((k-i)+\frac{1}{2}\big)x^2+d^2}~~~\forall~i\in [1,L]~,~\text{for}~\phi_i~,\nonumber\\
r_i &=& \sqrt{\big(k-i\big)x^2+d^2}~~~\forall~i\in [1,L-1]~,~\text{for}~\phi_{i,i+1}~, \nonumber\
\label{eq:displacement}
\end{eqnarray}
such that the Hamiltonian in Eq.~\eqref{eq:hamil} has the redshifted form as follows:
\begin{widetext}
\begin{equation}
H _\textrm{total}= J_x \sum_{i=1}^{L-1} \big[1+\phi_{i,i+1} (r_i)\big] \big(\hat{S}_i^x\hat{S}_{i+1}^{x} +\hat{S}_i^y\hat{S}_{i+1}^{y}\big) +J_z\sum_{i=1}^{L-1}  \big[1+\phi_{i,i+1} (r_i)\big] \hat{S}_i^z\hat{S}_{i+1}^{z} - \sum_{i=1}^L  \big[1+\phi_i(r_i)\big] B_i \hat{S}_i^z~.
\label{eq:hamil_red}
\end{equation}
\end{widetext}
From the form of the total Hamiltonian in Eq.~\eqref{eq:hamil_red}, one can see that the effect of gravity can equivalently be modelled by changing the local interaction constant $B_i$ and the nearest-neighbour coupling constants $J_s,~s\in\{x,y,z\}$. Quantum hardware platforms such as ion traps can be used to simulate many body systems such as spin chains \citep{Smith2019}. The coupling constants between quantum DOF in such platforms can be engineered with a good accuracy. Therefore, the effect of gravity and gravitational decoherence may be simulated on such platform in the near future to provide an experimental verification of our premise.

\begin{figure}
  \centering
    \includegraphics[width=1\columnwidth]{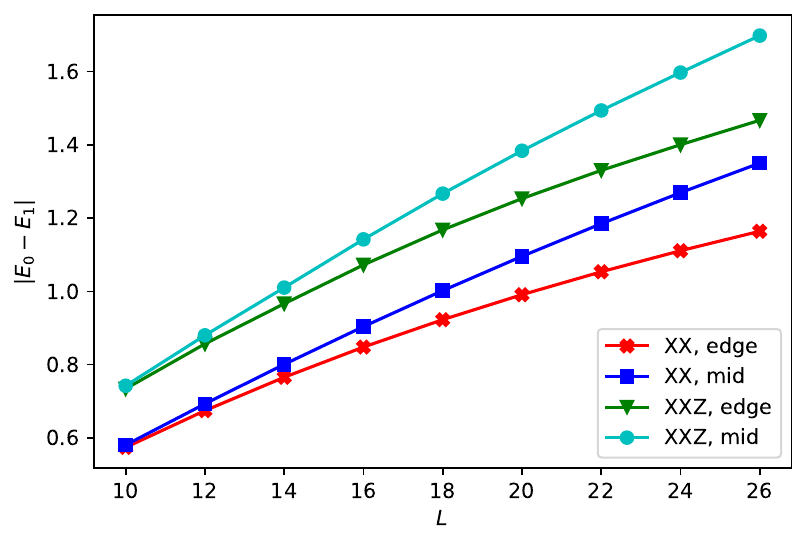}
   \caption{Change in the energy of the ground state ($|E_0-E_1|$) due to nearby gravitating particle with the length of the chain (L) for fixed $d=10$, $x=1$, and $GM=1$. The spin chain parameters are taken to be $J_x = J_z = B = 1$ for the XXZ chain and  $J_x = 1, J_z=B=0$ for the XX chain. $E_0$ is the ground state energy of chain with no gravitational particle nearby, $E_{1}$ refers to the ground state energy when the particle is at site $k_1$.}
  \label{fig:withL}
\end{figure}
\begin{figure*}
  \centering
    \includegraphics[trim={0.25cm 0cm 0.5cm 1cm},clip,width=2\columnwidth]{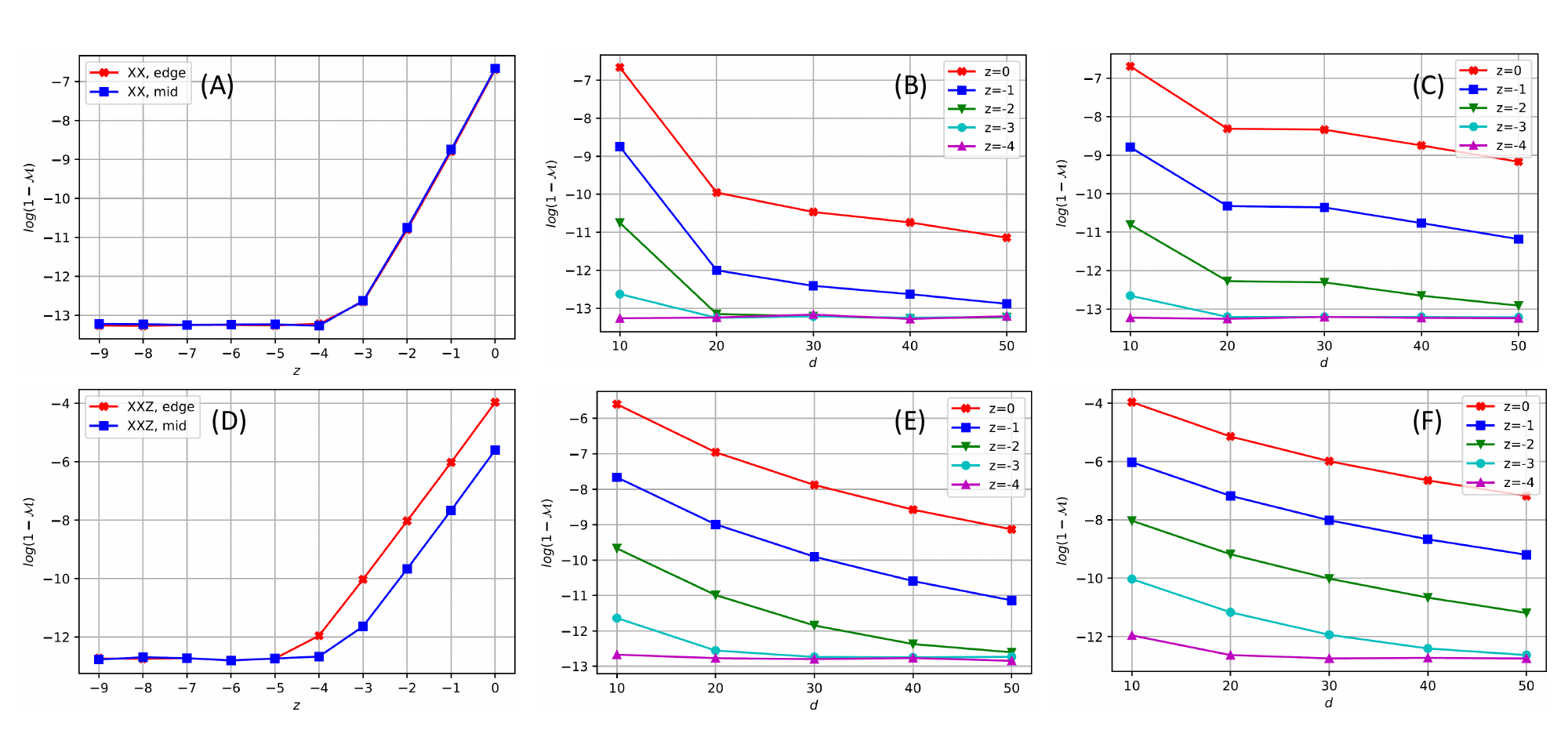}
 \caption{(A), and (D) show the variation of the distinguishability between the ground states of the spin chain corresponding to the sites $k_1$, $k_2$ with z which parameterizes the mass ($GM=10^z$) while keeping $d=10$ fixed for XX and XXZ chain respectively. Variation of the distinguishability between the ground states of the spin chain corresponding to the sites $k_1$ and $k_2$ of the spatial superposition of massive particle with the distance $d$ for different values of z for (B) middle configuration and (C) edge configuration of XX chain and (E) middle configuration and (F) edge configuration of XXZ chain. The XX spin chain parameters are taken to be $L=26$, $x=1$, $J_x = 1, J_z = B = 0$. The XXZ spin chain parameters are taken to be $L=26$, $x=1$, $J_x = J_z = B = 1$. }
  \label{fig:over_comp}
\end{figure*}

\begin{figure}
  \centering
    \includegraphics[width=1\columnwidth]{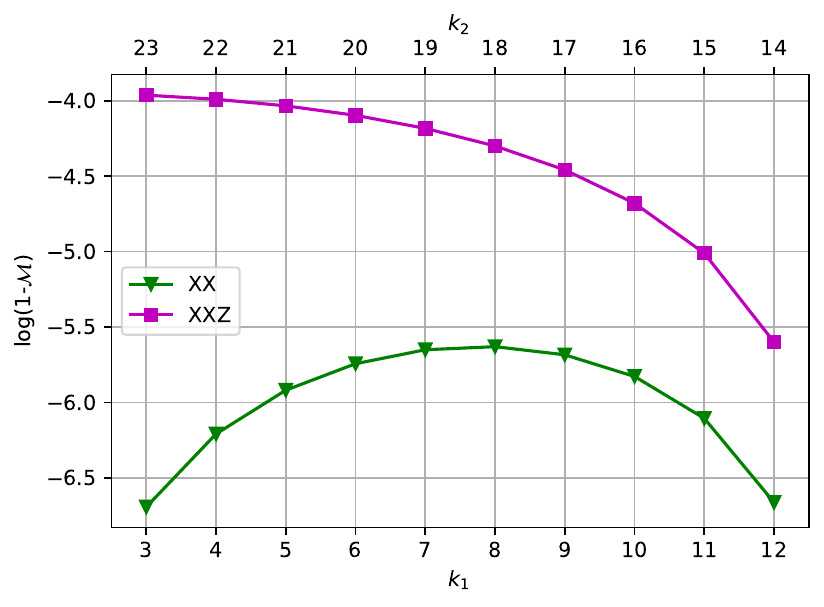}
    \caption{Variation of the distinguishability between the ground states of the spin chain corresponding to the sites $k_1$, $k_2$ with the site (k) for XX model, and XXZ model with z=0. The spin chain parameters are taken to be $L=26$, $x=1$, $J_x = J_z = B = 1$ for XXZ model and $L=26$, $x=1$, $J_x = 1, J_z = B = 0$. for XX model.}
  \label{fig:withsite}
\end{figure}

\subsection{Effect of a nearby massive particle on the ground state of a spin chain}

As a first step in quantifying the decoherence effect, we investigate the effect of a nearby massive particle on the ground state, and the ground state energy of XXZ and XX (by setting $J_z=B=0$ in Eq.~\eqref{eq:hamil}) spin chains. Calculations in this section are done numerically using the LANCZOS algorithm \cite{sandwik} in QUTIP \cite{qutip}.

We investigate the change in the ground state energy of the spin chain for 2 distinct positional configurations of the massive particle:
\begin{itemize}
    \item \textbf{Edge configuration :} Particle is placed near the edges ($k_1=3$, $k_2=L-3$ in Eq.~\eqref{eq:displacement}) 
    \item \textbf{Middle configuration :} Particle is placed approximately in the middle of the spin chain ($k_1=L/2-1$, $k_2=L/2+1$ in Eq.~\eqref{eq:displacement})
\end{itemize}

We denote $E_0$ as the ground state energy of the chain in
the absence of the massive particle, while $E_{1} (E_2)$ is the ground state energy when the particle is at site $k_1$ ($k_2$), see also Fig.~\ref{fig:grav}. The results are summarised
in Fig.~\ref{fig:withL}. The presented numerical analysis establishes
that the position of the massive particle influences
the ground state of the spin chain and thus also its
ground state energy. We note that, for both configurations
and both types of chains considered, the difference
$|E_2-E_1|$ is virtually zero as expected from these symmetric
(in space) models.

Note that in the continuum limit, $x\rightarrow 0$, the distance $|\tilde{r}_1-\tilde{r}_2|$ remains fixed, so the number of sites between the two locations scales as $k_2-k_1 \propto \frac{|\tilde{r}_1-\tilde{r}_2|}{x} \propto |\tilde{r}_1-\tilde{r}_2| \frac{L}{S}$, where $S$ is the size of the box in which the field is quantized (which also stays constant as we take the continuum limit). As both $L$, and $S$ approach $\infty$, the effect  remains dependent on a finite quantity, and therefore, does not vanish in the continuum limit.

\subsection{Effect of superposition of a massive particle on a
spin chain and decoherence}

As mentioned before, we will use $\mathcal{M}$ corresponding to different positions of the particle ($k_1$, $k_2$) to quantify the coherence of the spatial superposition which is also equivalent to the interferometric visibility in the \textit{Gedanken} experiment in Fig.~\ref{fig:ged_chain}. The decrease in $\mathcal{M}$ from $1$ is a sign of growing entanglement between the massive particle and the chain, which essentially leads to decoherence. Therefore, we use the distinguishability: $1-\mathcal{M}$ as a figure of merit for decoherence. In Fig.~\ref{fig:over_comp}, we show that the decoherence increases with the mass of the particle in superposition 
We parameterize the mass with $z$, where $GM=10^z$ and plot $\log{(1-\mathcal{M})}$ vs $z$ for both middle and edge configurations, in panels (A) for XX chain, and (D) for XXZ chain;  while keeping $d=10$ fixed. While the behaviour is the same for both the configurations in the XX chain, there is a difference, especially at higher $z$ for the XXZ chain. It is seen  from Fig.~\ref{fig:over_comp} (D) that there is a larger decoherence in the edge configuration than the middle configuration.

In Fig.~\ref{fig:over_comp} (B), (C) (XX chains), and (E), (F) (XXZ chains) we show the variation of decoherence with the distance between the spin chain and the mass for different values of the mass, i.e.~of $z$. As expected, the distinguishability between the ground states corresponding to sites $k_1$ and $k_2$ decreases as $d$ increases. This is equivalent to a larger value of the overlap $\mathcal{M}$ as shown for both configurations -- middle and edge in the (B), (E) and (C), (F) panels respectively. Since this is a direct indication of the influence of massive particle, this effect also depends on the mass. For higher values of $z$, the influence of the massive particle is greater even at larger values of $d$ which is reflected in a higher $1-\mathcal{M}$. Moreover, this effect also depends on the configuration of the spatial superposition of the massive particle which is clear on comparing the panels (B), (C) for XX chains and (E), (F) for XXZ chains.

To further analyse the site (k) dependence of $\mathcal{M}$, we consider $GM=1$, which indicates a very high mass for both XX and XXZ chains in Fig.~\ref{fig:withsite}. While the scale of decoherence differs considerably while the mass is changed, we aim to arrive at the highest decoherence achieved in the high mass limit. It is interesting to see that there is a qualitative as well as quantitative difference in the overlap resulting from XX and XXZ chains. As the symmetric sites are changed from near the edge to near the middle, while the distinguishability decreases monotonically (or equivalently the overlap increases) for the XXZ chain, it first increases and then decreases for the XX chain. One can expect that in the middle configuration i.e. for $k_1=L/2-1$, $k_2=L/2+1$, the overlap increases towards 1 as the states become less distinguishable. This inference is intuitive as the effect of the massive particle is most significant in the middle portion of the chain, and the slight asymmetry due to the mass being near the edge is absent. This is indeed observed from Fig.~\ref{fig:withsite} for XXZ chains, but the behaviour of XX chains remains elusive yet intriguing.

\subsection{Limiting case of black hole superposition}
\begin{figure}
  \begin{center}
    \includegraphics[width=1\columnwidth]{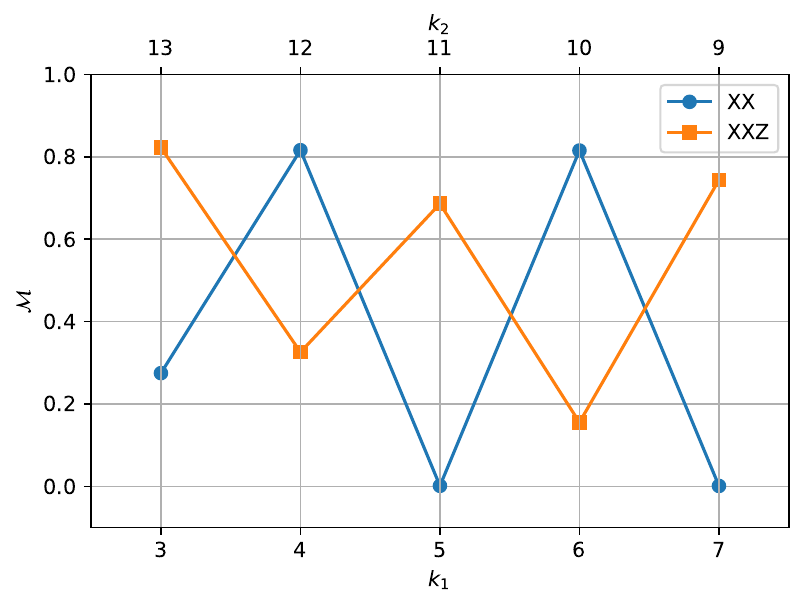}
  \end{center}
 \caption{ Overlap between the ground states of the redshifted Hamiltonians corresponding to the different sites $k_1$, and $k_2$ of the black hole. On the X-axis is the first site of the black hole. The second site is symmetrically located within the other half of the spin chain as shown on the top X-axis ($k_2=L-k_1$). $J_x = 1, J_z = B =0$ (XX model) and $J_x = J_z = B = 1$ (Heisenberg Model) while keeping $J_x=1, B=1$ for a spin chain of length $L=16$.}
  \label{fig:BHpl}
\end{figure}
We have shown that the increase in the distinguishability between the ground states of the spin chain corresponding to the different positions of the massive particle depends on the distance of the particle from the chain and its mass. The analysis in Fig.~\ref{fig:withsite} reveals that even at
very high mass of the particle, the change in magnitude of
distinguishability is not particularly high. This suggests
that in order to investigate the limits of the decoherence effect, we should consider the limit of high-mass of the
particle and small distance between the particle and the
chain. Therefore, here we consider a limiting case when the massive particle is a black hole with a Schwarzschild radius of $r_s=x/2$ occupying the space between $k$ and $k+1$ sites. Though not representative of a realistic scenario, this case theoretically allows for a (hypothetically) true constraint on the number of quantum DOF and the number of EQS in the spacetime region affected by the coherent superposition of a massive particle. 

We therefore investigate the limits of the effect by considering a black hole, as a matter of principle. For example, we investigate if one could get orthogonal ground states of the spin chain by taking a sufficiently large mass of the black hole. In effect, we expect the spin-chain model to break down when we take such massive particles. However, such a black hole model simply serves to put an absolute upper bound: in its regime of validity the model cannot produce  more decoherence than what we derive below in this limiting case.

In this scenario, the approximation given in Eq.~\eqref{eq:hamm} breaks down. Now, the redshifted Hamiltonian will be of the following form:
\begin{eqnarray}
H=\sum_i \sqrt{1-\frac{r_s}{r_i}}~h_i~,
\end{eqnarray}
where $r_i$ is taken from the centre of black hole and hence, for the $i^{th}$ site is as follows:
\begin{align}
r_i &=& |k-i+0.5|~x ~~~\forall~i\in [1,L]~,~\text{for}~\phi_i~,\nonumber\\
r_i &=& |k-i|~x ~~~\forall~i\in [1,L-1]~,~\text{for}~\phi_{i,i+1}~.
\label{eq:riBH}
\end{align}
The differential redshift factors $\phi_i$, $\phi_{i,i+1}$ lead to the final Hamiltonian of the same form as in Eq.~\eqref{eq:hamil_red}, but with a different $r_i$ dependence, as written in Eq.~\eqref{eq:riBH}.
For the purpose of calculating the ground state in the current case of black hole, we use exact diagonalization of the redshifted Hamiltonian.

We see in Fig.~\ref{fig:BHpl} that the overlap is considerably lower in value as compared to the previous cases with the nearby massive particle of considerable lesser mass than a black hole, see Fig.~\ref{fig:withsite} for comparison. This indicates that a finite number of spin chains may cause a complete decoherence of the Schwarzchild black hole in a spatial superposition located within the spin chain. In Fig.~\ref{fig:BHpl} we map the dependence of the overlap between the ground states corresponding to the black holes at different sites ($k_1$ and $k_2$) within the spin chain on the spin chain parameter $J_z$ using the representative cases of $J_z=1$ (Heisenberg Model) and $J_z=0$ (XX model). We see that both the site: $k_i$, and the value of $J_z$ control the overlap and in extension, the amount of decoherence caused by a single spin chain.

\subsection{Estimating the number of spin chains gravitationally interacting with a massive particle from the calculated coherence}
First, in order to arrive at the number of spin chains leading to the decoherence of a massive particle prepared in a spatial superposition, the \textit{Gedanken} experiment (see Fig.~\ref{fig:ged_chain}) could be realized experimentally with a mater-wave interferometer, where interference visibility can be obtained from successive runs
of the experiment. Very high visibility is achievable in such experiments, as demonstrated in recent works \cite{Asenbaum2017a, Fein2019}. Therefore, a substantial drop in visibility, of the order of $\frac{1}{10}$,  is certainly detectable.

Our results from the previous sections show that, if a single spin chain is taken, the resulting decoherence is
very small. This shall remain true irrespective of the
experimental setup, given the scaling of the effect with
the mass of the particle and a small mass-scale available
to table-top interference experiments \cite{Fein2019}. The drop in
the interference visibility purely due to the decoherence
mechanism discussed here for a single spin chain would
not be detectable within the limits of noise in any foreseeable
experiment. However, even if the effect is minuscule
with a single chain, a sufficient number of chains, can nevertheless, lead
to a detectable drop in the visibility. To formalise this idea, we denote an experimentally significant/measurable drop in visibility by $\alpha$. Thus we are interested in the case when $\mathcal{M} \rightarrow \alpha $. Considering $\alpha = e^{-1}$  to set the order of magnitude, we obtain the value for the number of independent systems $\mathcal{N}$ (see Section \ref{framework}) as follows:
\begin{eqnarray}
\mathcal{N} = \frac{-1}{\ln{ \mathcal{M}}}~.
\label{eq:constrain}
\end{eqnarray}
Since the overlap $\mathcal{M}$ is only slightly below one in all the cases, we approximate: $\ln{\mathcal{M}}=\ln{(1-\delta)}\approx-\delta$, with $\delta \ll 1$. In the best case, i.e. for the lowest overlap, we see that $\delta \approx 0.00016$ (for XXZ chain with the edge configuration of the spatial superposition of the massive particle as shown in Fig.~\ref{fig:withsite}). In this case, $\mathcal{N}=1/0.00016 \approx 6000$. On one hand, this shows that
a large number of spin chains is required to cause a significant
fall in the spatial coherence of a mass. On the
other hand, it also confirms our premise that coherence of the
spatial superposition of a massive particle can place an
upper bound (here $N=6000$) on the number of independent
systems with which the mass inevitably interacts.

It is important to note that we have considered a very high mass to calculate the overlap, and distinguishability for the above estimate. For example, the lowest overlap quoted above assumes a mass parameter: $z=0$, which translates to a mass in the \textit{Kilogram} range, for spin chains with the spacing in the \textit{Angstrom}  range. As discussed in the previous section, all other distances are taken on the scale fixed by the lattice spacing. The prospect of  bringing such a high mass in a spatial superposition looks unfeasible in the near future, as in the current record-setting experiments, e.g. in \citep{Fein2019}, the mass prepared in a spatial superposition is of the order of $10^{-23}$ kg.

 \section{Adiabatic condition in interference experiments}\label{discuss}
In our analysis, we have assumed that, for each position of the massive particle, the EQS ends up in the ground state of the corresponding red-shifted Hamiltonian. This requires that the position of the particle does not change too abruptly, so that the EQS stays at all times in the instantaneous ground state of the time-dependent Hamiltonian. For an interference experiment, this translates in the requirement that the superposition state is prepared sufficiently slowly.

We can evaluate the validity of this assumption for the case where the EQS is a fundamental field. As an estimate of the time scales involved, the time $T$ that it takes to prepare the superposition state should be long as compared to the gap $\Delta E$ between ground and first excited state of the field, $T \Delta E \gtrsim \hbar$. For a massive field, the gap $\Delta E$ is given by the mass $\mu$ of the corresponding fundamental particle, $\Delta E \sim \mu c^2$, giving us the estimate $T \gtrsim \frac{\hbar}{\mu c^2}$. As we are interested in particles that have not yet been discovered, we can take the mass scale to be larger than the heaviest known fundamental particle, the Higgs boson, $\mu \sim 10^{-25} \textrm{Kg}$. The corresponding time scale is of the order of $10^{-26} \textrm{s}$, which is extremely small compared to any laboratory procedure. For comparison, typical atom interferometers produce superposition states in times of the order of hundreds of milliseconds \cite{Asenbaum2017a}.

In short, the adiabatic assumption is valid as long as the interferometric procedure does not accidentally generate an excitation of the background field (i.e., create a particle), which seems very unlikely under ordinary conditions. Moreover, even if the adiabatic condition was not satisfied, we would expect the excitations generated by the experiment to radiate and carry which-way information, which would still result in a loss of visibility, i.e., decoherence.

\section{Conclusions} \label{conc}
In this work, we have incorporated the general relativistic redshift due to a localised source mass into a quantum Hamiltonian of an EQS. Because of the spatial extent of the EQS, the redshift depends on the distance between the massive particle and the local subsystems of the EQS. An EQS in
its ground state, at zero temperature, remains so when a
massive particle is brought close to it, as long as this is
done in an adiabatic manner. However, the exact ground state of the EQS Hamiltonian depends on the position of the massive particle due to the above redshift effect. This leads to entanglement between the position of the massive particle and the ground state of an EQS, which in turn leads to  decoherence of the massive particle in the position basis. This proposition is valid as long as the assumptions of universality of gravitational redshift and linearity of quantum mechanics hold correct simultaneously, which we presume to be true at least in the low energy limit. However, since the phenomenon of gravity-induced collapse still remains relevant in this regime (although unobserved), any predictions based on the spatial quantum superposition of massive particle -- which is the fundamental building block in our model -- belong to the realm of untested physics.

Since this phenomenon does not require any specific model of the EQS, it is universally valid for EQS with local interactions, including for quantum fields. The presence of many independent EQS leads to cumulative decoherence, with the total effect scaling exponentially with the number of such systems. We have proposed that by taking quantum fields as EQS, the decoherence of a massive particle could be used to estimate the number of fundamental quantum fields. Our work lays down a novel way to address a long standing problem in theoretical physics: How many fundamental fields are there? Despite that fact that we consider quantum states of active gravitational mass, our approach does not rely on explicit quantization of gravity, thus, our conclusions are compatible with any quantum gravity theory that may emerge in the future,  as long as such a theory has a `sensible' low-energy limit, respecting the universality of redshift and of quantum superpositions.

By incorporating the redshift due to a point-like mass
in an illustrative model of an EQS – a finite-length spin
chain – we analysed the resulting changes in the ground
state and in the associated energy. Importantly, if a massive
particle is placed symmetrically from the edges of the
chain, the difference between the ground state energies
of the chain corresponding to the different positions of
the particle is virtually zero; however, the corresponding
states are distinguishable.

The entanglement between the positional DOF of the massive particle and the ground state of the spin chain arising due to redshift leads to decoherence of the spatial superposition of the massive particle. This mechanism is similar to the entanglement between the position of a composite particle and its internal DOFs, due to the time dilation induced by a nearby mass on the internal dynamics \citep{Zych2011, pikovski2015}. In our formalism, without any reference to temporal dynamics, by the sheer dependence of the gravitational redshift (and in extension, of the ground state) on the position of the massive particle, we have shown that even a single spin chain can lead to a finite decoherence.

We have identified the coherence in the spatial superposition of the massive particle with the interferometric visibility in the \textit{Gedanken} experiment, which is equal to the overlap between the ground state of the spin chain corresponding to the different relative positions of the massive particle. In general, the distinguishability between the ground states decreases with the relative distance between the massive particle and the chain and increases with the mass of the particle. It also depends on the configuration of the spatial superposition (here symmetric with respect to the chain) and the type of spin chain.  It is useful to note that the mass of the particle which could lead to a measurable  decoherence quoted here would have to be very high. For example, the (relatively) high mass case that we have considered translates to the kilogram scale. Nevertheless, assuming the proposed model holds, the effect generally remains valid for small masses also (including that in present day matter-wave interferometers), though the bound, so obtained, would predict a very large number of EQS, and hence, may not be physically very meaningful for laboratory-controlled systems (such as spin chains).

By considering a scenario wherein a large number of spin chains are present in the vicinity of the massive particle, we ascertained that the total decoherence caused by them can be substantial, even for small masses. Assuming that the actual coherence observed in an experiment is decreased by a fraction from the maximum , we found that a finite -- though large -- number of spin chains would suffice to cause a significant decoherence.

In conclusion, we have introduced the idea of a differential redshift --  extending the well known paradigm of gravitational effects on quantum systems to a novel context of EQS. Further, we have identified a resulting decoherence effect for massive particles in spatial superpositions and  have illustrated it for a spin chain as a toy model for a one-dimensional EQS. We have proposed that the effect we found  extends to the continuum limit, including to relativistic quantum fields. If confirmed, our idea will provide a tool to estimate an upper bound on the total number of fundamental particles, based on current, low energy experiments.

\acknowledgements
M.Z and F.C.~acknowledge support through Australian Research Council (ARC) DECRA grants DE180101443 and DE170100712, and ARC Centre EQuS CE170100009. The authors acknowledge the traditional owners of the land on which the University of Queensland is situated, the Turrbal and Jagera people.

\bibliography{reff.bib}
\end{document}